\documentclass[12pt,preprint]{aastex}

\newcommand\lsim{\lower0.5ex\hbox{$\; \buildrel < \over \sim \;$}}

\shorttitle{Search for magnetic monopoles} \shortauthors{Zhang et
al.}

\begin{document}

\title{Search for Magnetic Monopoles in Magnetic Reconnection Regions}

\author{Jun Zhang\altaffilmark{}, Ting Li\altaffilmark{}}

\altaffiltext{}{Key Laboratory of Solar Activity, National
Astronomical Observatories, Chinese Academy of Sciences, Beijing
100012; University of Chinese Academy of Sciences, Beijing 100049,
China; zjun@nao.cas.cn; liting@nao.cas.cn}

\begin{abstract}

In order to satisfy the symmetry between electric and magnetic
fields in the source free Maxwell's equations, electric charges
might have magnetic counterparts: magnetic monopoles. Many methods
and techniques are proposed to search for the monopoles, but no
confirmed results have been obtained. Based on solar observations,
we know that magnetic reconnections take place during eruptive solar
activities. The magnetic fields can be broken at first and then
rejoined, implying that the fields are source-relevant at the broken
moment. It is speculated that the magnetic lines undergo outward
deflection movement during the broken moment, as the line tying
effect disappears and the magnetic tension triggers the movement.
The signal of the deflection is detected for the first time by EUV
and H${\alpha}$ observations in reconnection processes. We propose
that the monopoles appear in magnetic reconnection regions at first,
and then the annihilation of opposite polarity monopoles releases
energy and perhaps also produces particles. To detect the predict
monopoles, laboratory plasma experiments can be used to provide some
fundamental information.

\end{abstract}


\section{Introduction}

Following the predicted existence of monopoles from spontaneous
symmetry breaking mechanisms (Dirac 1931), searches have been
routinely made for monopoles produced at accelerators, in cosmic
rays, and bound in matter (Nakamura \& Particle Data Group 2010).
The main strategy to search for monopoles is that monopoles will
interact with their pass through matter (Fairbairn et al. 2007). It
is suggested that monopoles with Dirac charge would typically lose
energy at a rate which is thousands times larger than that expected
from particles with the elementary electric charge. Consequently,
scintillators, gas chambers and nuclear track detectors have been
used in cosmic ray and collider experiments (Price et al. 1975;
Cabrera 1982). Many efforts have been made to seek monopoles in moon
rock, meteorites and sea water (Kovalik \& Kirschvink 1986; Jeon \&
Longo 1995). Furthermore, a range of experiments about the induced
nucleon decay are also employed to detect the monopoles (Kajita et
al. 1985; Bartelt et al. 1987; Becker-Szendy et al. 1994; Balkanov
et al. 1998; Ambrosio et al. 2002). Collider experiments about lower
energy (Carrigan et al. 1973; Aubert et al. 1983) and high energy
(Kalbfleisch et al. 2004; Abulencia et al. 2006) in hadron-hadron
collisions have been employed to search for the suggested monopoles.

In solar and stellar coronae, magnetic reconnection is an essential
physical process (Cassak et al. 2008; Zhang et al. 2013). Vast
theoretical studies of magnetic reconnection have been done to
explain flares (Sturrock 1966; Hirayama 1974; Kopp \& Pneuman 1976)
and filament eruptions (Shibata 1999; Antiochos et al. 1999; Lin \&
Forbes 2000). Magnetic reconnection occurs at an X-point where
anti-parallel magnetic field lines converge and reconnect, then the
magnetic energy is released and converted to other forms of energy,
e.g. radiation, energetic particle acceleration, and kinetic energy
of plasma (Priest \& Forbes 2000; Priest 2014). Many signatures of
magnetic reconnection have been observed in different kinds of solar
eruptions, e.g., flares and coronal mass ejections, and as well as
solar winds (Tsuneta et al. 1992; Masuda et al. 1994; Shibata et al.
1995; Yokoyama et al. 2001; Innes et al. 2003; Sui \& Holman 2003;
Asai et al. 2004; Lin et al. 2005; Gosling et al. 2007; Li \& Zhang
2009; Zhang et al. 2013). In the magnetosphere, magnetic
reconnection are also revealed by in situ measurements (Phan et al.
2007; Mozer et al. 2002; Dunlop et al. 2011). Moreover, many
experiments which are dedicated to magnetic reconnection have been
carried out in laboratories under controlled conditions (Bratenahl
\& Yeates 1970; Yamada et al. 1997).

In this paper, we report new properties of magnetic reconnections
and propose that the magnetic monopoles can be searched for in
magnetic reconnection regions. We describe the observational data in
Section 2. In Section 3, we present the results. The conclusions and
a brief discussion are displayed in Section 4.

\section{Observations}

The New Vacuum Solar Telescope (NVST; Liu et al. 2014) observe the
Sun with high temporal and spatial resolutions. We mainly use the
H${\alpha}$ line from the NVST to study the dynamic evolution of
small-scale magnetic reconnection. Moreover, the Atmospheric Imaging
Assembly (AIA; Lemen et al. 2012) multi-wavelength observations and
the Helioseismic and Magnetic Imager (HMI; Scherrer et al. 2012;
Schou et al. 2012) line-of-sight magnetograms from the Solar
Dynamics Observatory (SDO; Pesnell et al. 2012) are also used.
\emph{SDO}/AIA observes the full disk of the Sun in 10 wavelengths
with a pixel size of 0$\arcsec$.6 and a cadence of 12 s. These data
reveal the solar atmospheric temperatures from $\sim$5000 K to
$\sim$20 MK. The \emph{SDO}/HMI records the line-of-sight (LOS)
magnetic field with a cadence of 45 s and a spatial sampling of
0$\arcsec$.5 pixel$^{-1}$.

On 27 January 2012, two sets of EUV loops observed by AIA appeared
above the western solar limb as seen from the Earth (Sun et al.
2015). AIA 171 {\AA} and 94 {\AA} images from $\sim$ 00:00 to 07:00
UT are adopted to research the process of magnetic reconnection. On
2014 February 3, the NVST observed the AR 11967 with a field of view
(FOV) of 151$''$ ${\times}$ 151$''$, and the H${\alpha}$ 6562.8
{\AA} images are adopted in this study from 05:49:52 UT to 09:10:01
UT, with a cadence of 12 s and a spatial sampling of 0.$''$163 per
pixel. The calibration, correction and speckle masking (Weigelt
1977; Lohmann et al. 1983) reconstruction of these H${\alpha}$ data,
as well as the co-alignment between these H${\alpha}$ data and SDO
images are described by Yang et al. (2015).

\section{Results}

The concept of magnetic reconnection was proposed long time ago, and
many observations and theoretic models have been displayed and put
forward. However, the most important question (aspect) that whether
magnetic field breaks or not during reconnection has always been
omitted. The advance of the solar observations help us to check the
details of magnetic reconnection. This work tracks the following
idea. If magnetic field breaks during reconnection, the broken field
will undergo a special outward movement, due to the disappearance of
the line tying effect and magnetic tension causes the movement of
the broken field. Examining this movement will provide new clues
about the essentials of reconnection, such as the magnetic fields
are source-relevant, and monopoles or equivalent monopoles appear at
the broken points.

\subsection{Observations of magnetic reconnection in solar atmosphere}

Recently, both the ground-based and space-borne observations have
provided a mass of magnetic reconnection events in solar atmosphere.
Two events are chosen to display the properties of magnetic
reconnection. The first event occurred on 27 January 2012. From
$\sim$ 00:00 to 03:00 UT, two sets of cool loops (171 {\AA}
passband, corresponding to a temperature of $\sim$0.6 MK) which were
above the western solar limb (Fig. 1a) moved towards each other and
formed an X-shaped structure near 03:00 UT (Fig. 1b). Following the
disappearance of the cool loops, a hot region ($\sim$7 MK; detected
in the channel of AIA 94 {\AA}) immediately appeared near the
X-shaped structure, implying the initial heating of a solar flare
(red in Fig. 1c). About one hour later, post flare loops were
detected (Fig. 1d).

We speculate that the broken fields will undergo a special movement,
so we focus on the evolution of EUV loops while the two sets of
loops approach. Indeed, the loops at the both sides of the ``Void"
space (denoted by arrows in Fig. 1c) deflect outward, and this
deflection has never been reported before. At the beginning of the
reconnection the deflection (Fig. 2b and 2c) speed is small, i.e.,
1.5 Mm in 13 min, or 1.1 km s$^{-1}$. At the later phase of the
reconnection, the deflection (Fig. 2e and 2f) speed is 2.2 km
s$^{-1}$, twice as large as that at the beginning. An animation
(movie1.mpeg) to display this deflection is available in the online
journal.

The multi-wavelength observations from the AIA with the high spatial
and temporal resolution successfully detect evidences for
reconnection including plasma inflows, heating close to the
reconnection site, and outward deflection relative to the broken
fields. To quantitatively analyze the inflows and deflection, we
select three slices in the 171 and 94 {\AA} composite images (``AB",
``CD" and ``EF" in Fig. 1a). The stack plots (Fig. 3) clearly show
that the bilateral cool loops (cyan) keep moving towards the
reconnection region. Once the visible innermost loops come into
contact at $\sim$ 03:00 UT (blue dashed lines in Fig. 3a and 3c),
hot plasma (red, Fig. 3b and 3c) appears at the reconnection site.
The speeds of the inflows vary from 1.5 to 4.1 km s$^{-1}$ (Fig. 3a
and 3b). In addition, the average outward deflection speed of the
loops at the upper side of the ``Void" space is about 1.4 km
s$^{-1}$.

The second reconnection event is displayed in Fig. 4. Sequence of
H$\alpha$ images show the magnetic reconnection between two sets of
small-scale loops ``L1" and ``L2", as shown in panel (a). Prior to
the initiation of rapid reconnection, the two sets of loops ``L1"
and ``L2" were moving towards each other and then interacted. At
07:18:52 UT, both loops ``L1" and ``L2" were apparently broken (as
denoted in panel (b)). Meanwhile, two broken loops (``B$_{L1}$" and
``B$_{L2}$" in panel (b)) which were respectively related to ``L1"
and ``L2" appeared, but the curvatures of ``B$_{L1}$" and
``B$_{L2}$" were more different from that of ``L1" and ``L2". Then
``B$_{L1}$" and ``B$_{L2}$" connected to form a loop ``L4", and
loops ``L1" and ``L2" disappeared (panel (c)). To better exhibit
this reconnection process, an animation (movie2.mpeg) is available
in the online journal. At the reconnection region and its proximity,
the brightenings in the H${\alpha}$ images (panels (b) and (c)) and
the 171 and 94 {\AA} composite images (panel (d)) can be found. As
displayed in panel (e), all the three light curves in the
reconnection region (blue window in panel (a)) reach the peak
simultaneously, i.e., $\sim$07:18:50 UT. To better exhibit this
reconnection process, an animation (movie2.mpeg) is available in the
online journal.

\subsection{Magnetic monopoles appearing in magnetic reconnection regions}

%\subsubsection{First step of magnetic reconnection: magnetic fields are broken}

Traditional reconnection models presented that opposite magnetic
fields form a current sheet while the fields approach each other.
Magnetic energy converts into heat and kinetic energy by Ohmic
dissipation in the current sheet which locates in a tiny diffusion
region. But the joint between the fields and the current sheet is
always omitted, and nobody knows the physical properties of the
joint. Although Dungey (1953) was the first to suggest that ``lines
of force can be broken and rejoined", in the past decades and at
present time, almost no researcher have considered that magnetic
field lines must be broken firstly, if the lines are involved in
reconnection. We suggest that while the field lines break, opposite
polarity monopoles at the two broken points appear. Figure 5
displays a series of schematic drawings which illustrate the
magnetic reconnection process. The green arrows in panel (a) denote
the convergence of two sets of loops (``L1" and ``L2", ``L3" and
``L4"), and the vertical red structure (panels (b) and (c))
represents the current sheet (CS). At the joint (denoted by a blue
point in panel (c)) between CS and L2, loop ``L2" breaks. The
corresponding magnetic field $\overrightarrow{B2}$ breaks also and
its direction points to the joint. So $\overrightarrow{B2}$ is
source-relevant at this moment and negative monopoles $Q_{-m}$
appears at the joint. Similarly, ``L3" and its corresponding
magnetic field $\overrightarrow{B3}$ break at another joint (red
point in panel (c)). Positive monopoles $Q_{m}$ appear at this
joint, as $\overrightarrow{B3}$ points outward from the joint.
Meanwhile, the curvatures of ``L2" and ``L3" in panel (b) change to
that in panel (c). The main reason is that the line tying effect
does not work during the broken process, and magnetic tension
triggers the change of the curvatures. The Gauss's law for magnetism
around the two red circles in panel (c) can be written respectively
as

${\nabla}{\cdot}\overrightarrow{B2}={\alpha}Q_{-m}$   (1) \\

${\nabla}{\cdot}\overrightarrow{B3}={\alpha}Q_{m}$    (2) \\

Where ${\alpha}$ is a coefficient. We suggest that the monopoles are
instable, they should be annihilated in a short time. It is possible
that ``L2" and ``L3" (panel (c)) connects to form a new loop (panel
(d)). The annihilation of opposite polarity monopoles releases
energy ($En$) and perhaps produces particle ($P$). This process can
be expressed as

$Q_{-m}+Q_{m}{\longrightarrow}En+P$    (3) \\

\section{Conclusions and Discussion}

Based on both the ground-based and space-borne observations, we
report two magnetic reconnection events in solar atmosphere.
Assuming that magnetic fields can be broken at first and then
rejoined during reconnection process, we suggest that the magnetic
lines undergo outward deflection movement during the broken moment,
as the line tying effect does not work and the magnetic tension
triggers the movement. The signal of this movement is detected for
the first time by EUV and H${\alpha}$ observations in the two
reconnection events. To satisfy the Gauss's law for magnetism, we
propose that the monopoles appear in magnetic reconnection regions.
The annihilation of opposite polarity monopoles releases energy and
perhaps also produces particles, then new loop forms. It is
speculated that the monopoles can be searched for in magnetic
reconnection region, if the monopoles exist indeed in nature.

The common features of most reconnection theories include the
changes of magnetic topologies and the release of magnetic energy
(Parker 1957; Sweet 1958; Petschek 1964; see the review by Yamada et
al. 2010). The topology changes are mainly the break of inflowing
anti-parallel loops and the formation of new loops. When the loops
reconnect in the diffusion region, magnetic energy is released, thus
heating the plasma. In addition, the reconnected field lines near
the X-point are sharply bent and the magnetic tension force also
impacts the plasma to increase the kinetic energy. Therefore, the
plasma is brightened and expelled. However, it is very few to take
into account the break of magnetic field and the succedent matter of
the break. In this work, we forecast at first the deflection
movement, and then we find the signal of the movement in the
reconnection events.

Since the introduction of magnetic monopoles by Dirac (1931), they
are compulsory in many formulations of Grand Unified Theories (GUT,
Polyakov 1974; 't Hooft 1974). Although GUT-scale monopoles are
commonly believed to be extremely heavy (10$^{17}$ GeV), there are
other mechanisms resulting in production of much lighter monopoles
after inflation. Kephart and Shafi (2001) proposed that monopoles
with a magnetic charge of a few Dirac units and masses in the range
10$^{7}$$\sim$10$^{13}$ GeV could be occurred in symmetry-breaking
events. Furthermore, there is a speculation that these lighter
monopoles are possible ultra-high energy cosmic rays, so monopoles
can be searched for in cosmic radiation. Based on both measurements
and estimates of cosmic magnetic fields, it is suggested that they
could accelerate monopoles lighter than 10$^{14}$ eV to relativistic
velocities (Beck et al. 1996; Ryu et al. 1998).

It is a fact that no traditional monopoles have been verifiably
detected (Yao et al. 2006). Although there are some reports of
magnetic monopole detections (Price et al. 1975; Cabrera 1982;
Caplin et al. 1986), these reports have often been challenged by the
original authors themselves (Price et al. 1978; Huber et al. 1990).
In this paper, we propose a new idea: search for magnetic monopoles
in magnetic reconnection regions. Both remote observations and local
detections can be employed to search for monopoles in reconnection
regions which are in solar atmosphere and in laboratories,
respectively.

\acknowledgments {This work is supported by the National Natural
Science Foundations of China (11533008, 11221063, 11203037, 11303049
and 11303050) and the Strategic Priority Research Program$-$The
Emergence of Cosmological Structures of the Chinese Academy of
Sciences, Grant No. XDB09000000. The data are used courtesy of
NASA/\emph{SDO} and NVST science teams. }

{}

\clearpage

\begin{figure}
\includegraphics
[bb=47 237 540 590,clip,angle=0,width=\textwidth]{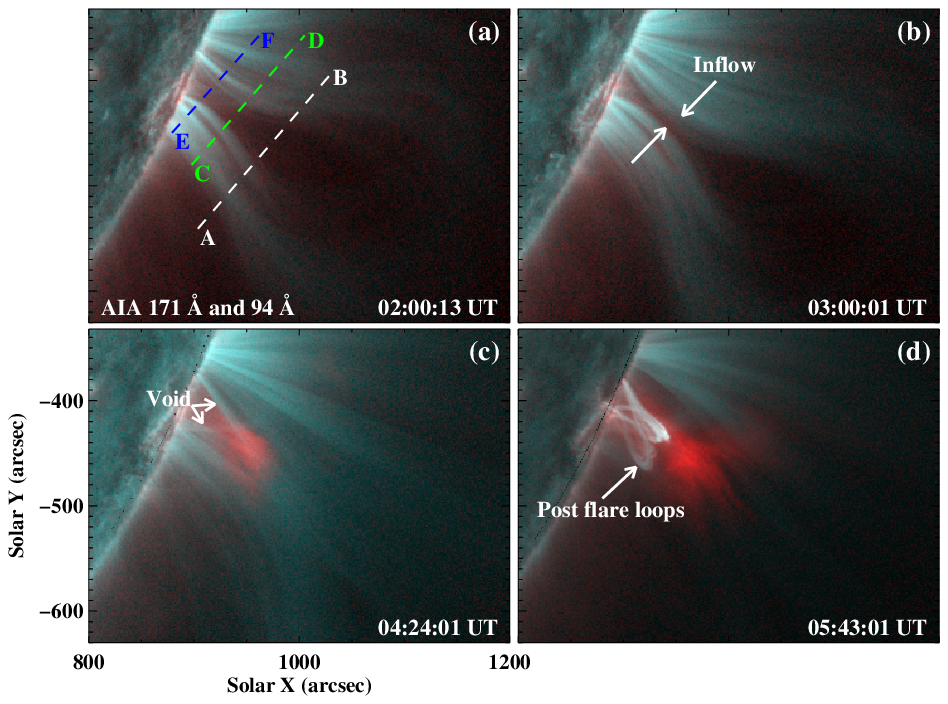}
\caption{ Composite images of the AIA 94 {\AA} (red) and 171 {\AA}
(cyan) passbands showing the reconnection event on 27 January 2012.
Lines ``AB", ``CD" and ``EF" in panel (a) denote the positions that
are used to obtain the stack plots displayed in Figure 3.
\label{fig1}}
\end{figure}

\clearpage

\begin{figure}
\includegraphics
[bb=47 237 540 590,clip,angle=0,width=\textwidth]{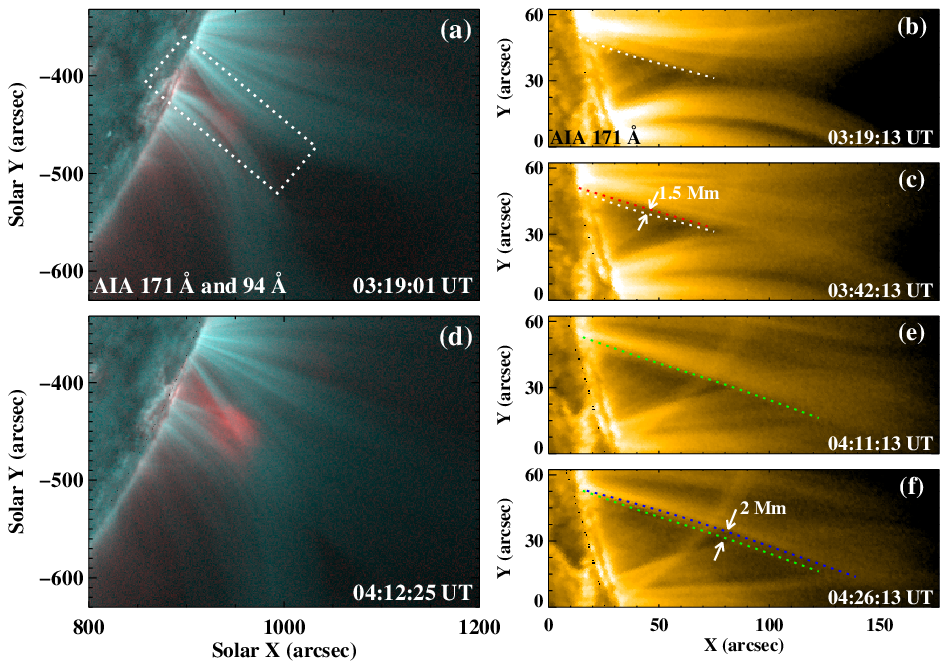}
\caption{Panels (a) and (d): two composite images of the AIA 94
{\AA} (red) and 171 {\AA} (cyan) passbands showing the approach (a)
and reconnection (d) of two sets of coronal loops. The window in
panel (a) outlines the field-of-view of the 171 {\AA} images in
panels (b), (c), (e) and (f). Panels (b) and (c): outward deflection
of the upper set of loops at the beginning of the reconnection.
White (red) lines denote the inner boundary of the loops at 03:19:13
UT (03:42:13 UT). Panels (e) and (f): similar to panels (b) and (c),
outward deflection of the upper set of loops at the later phase of
the reconnection. Green (blue) lines denote the inner boundary of
the loops at 04:11:13 UT (04:26:13 UT). An animation (movie1.mpeg)
of this figure is available in the online journal. \label{fig2}}
\end{figure}

\clearpage

\begin{figure}
\includegraphics
[bb=161 297 409 530,clip,angle=0,width=0.8\textwidth]{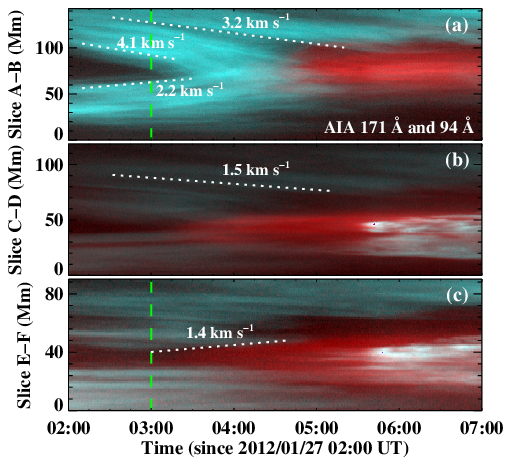}
\caption{Temporal evolution of plasma inflows (see the dotted lines
in panels (a) and (b)) and deflecting loops (denoted by the dotted
line in panel (c)) during the reconnection. \label{fig3}}
\end{figure}

\clearpage

\begin{figure}
\centering
\includegraphics
[bb=47 237 551 596,clip,angle=0,width=\textwidth]{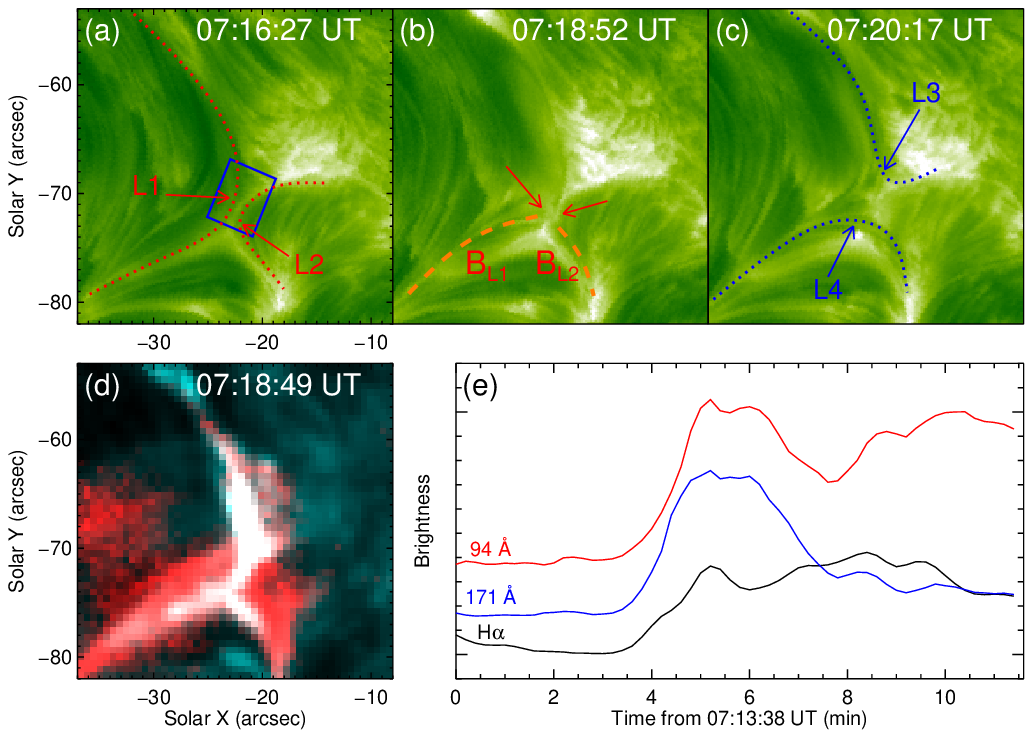}
\caption{Panels (a)-(c): time sequence of H$\alpha$ images showing
the reconnection process between two sets of small-scale loops.
Arrows ``L1" and ``L2" point to the loops before reconnection, and
arrows ``L3" and ``L4" denote the newly formed loops after
reconnection. Two arrows in panel (b) denote two broken points, and
the dashed curves B$_{L1}$ and B$_{L2}$ indicate two broken loops.
Panel (d): composite image of the AIA 94 {\AA} (red) and 171 {\AA}
(cyan) passbands showing the brightening at reconnection moment.
Panel (e): light curves in the H$\alpha$, 171 {\AA}, and 94 {\AA}
lines obtained from the area within the blue window in panel (a). An
animation (movie2.mpeg) for this reconnection event is available in
the online journal. \label{fig}}
\end{figure}

\clearpage

\begin{figure}
\centering
\includegraphics
[bb=146 178 451 689,clip,angle=0,width=0.6\textwidth]{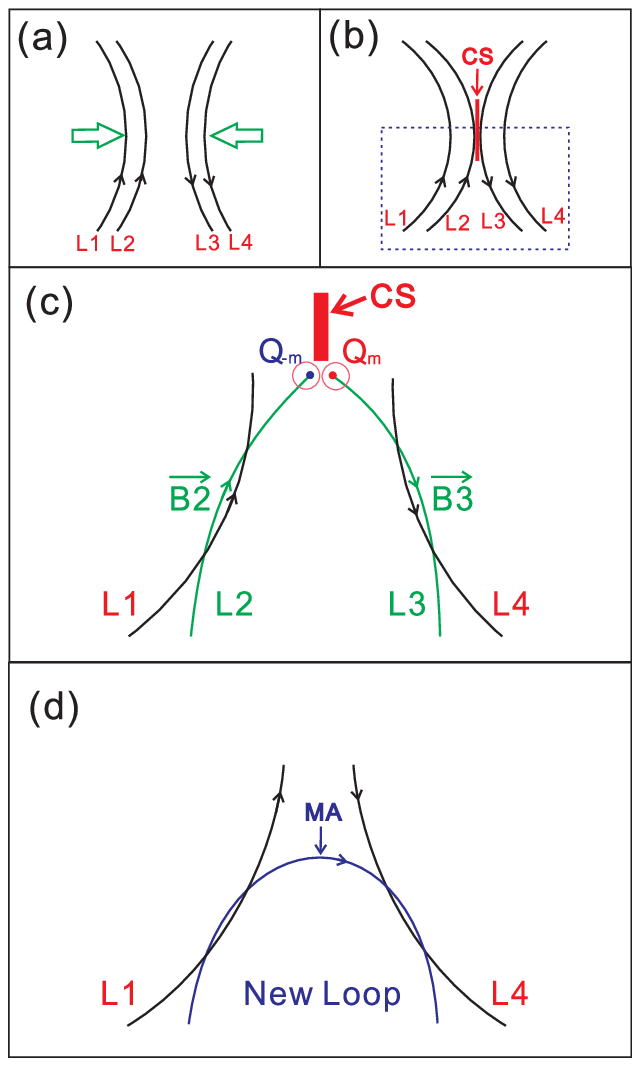}
\caption{Schematic drawings illustrating the magnetic reconnection
process. The green arrows in panel (a) denote the convergence of two
sets of loops (``L1" and ``L2", ``L3" and ``L4"). The vertical red
structure represents the current sheet (CS), and the blue dotted
rectangle outlines the field-of-view of panels (c)-(d). The green
curves in panel (c) represent the broken loops of ``L2" and ``L3",
$\protect\overrightarrow{B2}$ and $\protect\overrightarrow{B3}$ are
the magnetic fields of ``L2" and ``L3", respectively. Q$_{m}$ and
Q$_{-m}$ are the positive and negative monopoles which appear near
the ends of the broken loops. The blue curve in panel (d) indicates
the newly formed loop, after the monopoles annihilation (MA).
\label{fig}}
\end{figure}

\clearpage

\begin{figure}
\includegraphics[bb=46 255 515 789,clip,angle=0,width=0.7\textwidth]{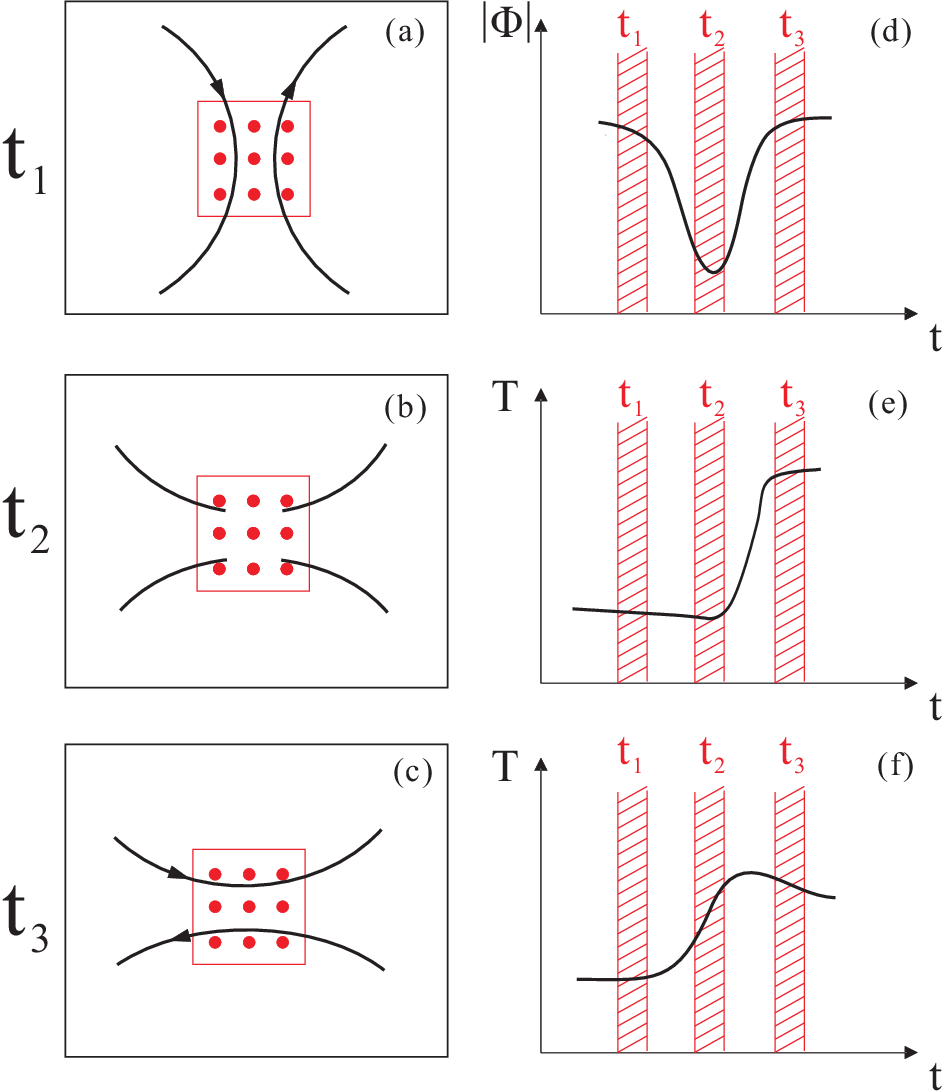}
% Here is how to import EPS art
\caption{\label{fig:epsart}Panels (a)-(c): schematic drawings
illustrating the magnetic reconnection process which will be
detected by laboratory plasma experiments. Red windows and red dots
in the windows represent probe array to measure magnetic flux
$\mid\Phi\mid$ and ion temperature ``T". Panels (d)-(e): possible
changes of magnetic flux $\mid\Phi\mid$ (d) and ion temperature ``T"
(e) during the reconnection process (t$_{1}$-t$_{3}$ in panels
(a)-(c)). Magnetic fields break and magnetic monopoles appear at
t$_{2}$ moment, then monopoles annihilation (t$_{3}$) will transfer
the magnetic energy to thermal energy and increase the ion
temperature. Panel (f): magnetic energy transforming into the ion
thermal energy by current sheet dissipation during the interval
t$_{1}$ and t$_{3}$, as suggested by traditional magnetic
reconnection models. \label{fig}}
\end{figure}

\end{document}